  \documentclass[superscriptaddress,preprint,showpacs,amsmath,amssymb,floatfix]{revtex4-1}

\usepackage{amsmath}    
\usepackage{amsfonts}   
\usepackage{amssymb}
\usepackage{graphicx}   
\usepackage{subfigure}
\usepackage{overpic}
\usepackage{bbm}

\newcommand{\beq}{\begin{equation}}
\newcommand{\eeq}{\end{equation}}

\newcommand{\bfm}[1]{\mbox{\boldmath${#1}$}}

\begin{document}

\title{Nonlinear Kinetics on Lattices based on the Kinetic Interaction Principle}
\author{Giorgio Kaniadakis}
 \email{giorgio.kaniadakis@polito.it}
 \affiliation{Department of Applied Science and Technology, Politecnico di Torino,
Corso Duca degli Abruzzi 24, 10129 Torino, Italy}
\author{Dionissios T. Hristopulos}
 \email{dionisi@mred.tuc.gr}   
 \affiliation{School of Mineral Resources Engineering, Technical University of Crete,
Chania 73100, Greece}
\affiliation{Telecommunication Systems Institute, Chania 73100, Greece}

\date{\today}

\begin{abstract}
Master equations define the dynamics that govern  the time evolution of various physical processes on  lattices. In the continuum limit, master equations lead to Fokker-Planck partial differential equations that represent the dynamics of physical systems in  continuous spaces. Over the last few decades,  nonlinear Fokker-Planck equations have become very popular in condensed matter physics and in statistical physics. Numerical solutions of these equations require the use of discretization schemes. However,
 the discrete evolution equation obtained by the discretization of a Fokker-Planck partial differential equation depends on the specific discretization scheme. In general, the  discretized form is different from the master equation that has generated the respective Fokker-Planck equation in the continuum limit. Therefore, the knowledge of the master equation associated with a given Fokker-Planck equation is extremely important for the correct numerical integration of the latter, since it provides a unique, physically motivated  discretization scheme. This paper shows that the Kinetic Interaction Principle (KIP) that governs the particle kinetics of many body systems, introduced in [G. Kaniadakis, Physica A {\bf 296}, 405 (2001)], univocally defines a very simple master equation that in the continuum limit yields the nonlinear Fokker-Planck equation in its most general form.
\end{abstract}



\maketitle


\section{Introduction}

Nonlinear kinetics in the coordinate or velocity space or in the phase space has been used systematically over the last few decades. A popular approach involves studying the temporal evolution of a particle system while it interacts with a medium in thermodynamic equilibrium. In ref.~\cite{PhysicA2001}, the most general nonlinear kinetics in  phase space was considered in the limit of the diffusive approximation. In this limit, the interaction between the particles and the medium is viewed as a diffusive process in the velocity space. It was shown that the \emph{particle probability density function} $f=f(t,\mbox{\boldmath$r$},\mbox{\boldmath$v$})$ (henceforward particle density function) obeys the following evolution equation
\begin{eqnarray}
\frac{\partial f}{\partial t}+
\frac{1}{m}\frac{\partial U}{\partial
\mbox{\boldmath $v$}}\frac{\partial f}{\partial \mbox{\boldmath
$r$}}-\frac{1}{m}\frac{\partial V}{\partial
\mbox{\boldmath $r$}}\frac{\partial f}{\partial \mbox{\boldmath
$v$}}= \frac{\partial}{\partial \mbox{\boldmath$v$}}
\left \{ D a(f)\, b(f)\frac{\partial}{\partial
\mbox{\boldmath$v$}} \bigg [ \beta V+ \beta U  \!+\! \ln \frac{a(f)}{b(f)} \bigg ] \! \right \}, \ \ \  \ \ \ \   \label{I1}
\end{eqnarray}
where $D=D(\bfm{r},\bfm{v})$ is the diffusion coefficient, while $V=V(\bfm{r})$ and $U=U(\bfm{v})$ represent the external potential and the particle kinetic energy respectively. In addition, $\partial f/ \partial \bfm{r}$, $\partial f/ \partial \bfm{v}$, denote  the gradients of the particle density with respect to  velocity and space, and $m$ is the particle mass.
The constant $\beta$ is given by $ \beta = 1 / k_{\mathrm{B}} T $  where $k_{\mathrm{B}}$ is the \emph{Boltzmann constant} and $T$ the temperature of the system.

 The above equation can be viewed as a \emph{generalized Kramers equation} that describes a nonlinear kinetics in the phase space. The nonlinearity is due to the presence of the two arbitrary functions $a(f)$ and $b(f)$ that depend on the particle density function $f$. In the case where $a(f)=f$ and $b(f)=1$, Eq.~(\ref{I1}) reduces to the ordinary Kramers equation that describes the standard linear kinetics, e.g.~\cite{Risken84}.

Eq.~(\ref{I1}) has been  employed to study the kinetic foundations of nonlinear statistical systems governed by $\kappa$-entropy. Generalized statistical mechanics, based on $\kappa$-entropy \cite{PhysicA2001,KQSphysA2003,PRE2017}, preserves the main features of ordinary Boltzmann-Gibbs statistical mechanics. For this reason, it has attracted the interest of many researchers over the last 16 years, who have studied its foundations and mathematical aspects \cite{Silva06A,Naudts1,Topsoe,Tempesta2011,Scarfone2013,SouzaPLA2014,Scarfone1,Scarfone2}, the underlying thermodynamics \cite{Wada1,ScarfoneWada,ScarfoneWadaJPA2014,Bento3lawThermod,WadaMatsuzoeScarfone2015}, and specific applications of the theory in various scientific and engineering fields. A non-exhaustive list of application areas includes quantum statistics \cite{Santos2011a,Planck,Lourek2},  quantum entanglement  \cite{Ourabah,OrabahPhyscripta},  plasma physics \cite{Lourek,Gougam,Chen,Landau2017,Qualitative2017},  nuclear fission \cite{NuclearEnergy2017},  astrophysics \cite{Carvalho,Carvalho2,Carvalho2010,Cure,AbreuEPL,AbreuIJMPA,ChinesePL}, geomechanics \cite{Oreste2017},  genomics \cite{SouzaEPL2014},  complex networks \cite{Macedo,Stella},  economy \cite{Clementi2009,Bertotti,Modanese,Bertotti2017} and finance \cite{Trivellato2012,Trivellato2013,Tapiero,Trivellato2017}.

It is important to note that the right hand side of Eq. (\ref{I1}), which introduces nonlinear effects into kinetics, describes an unconventional diffusion process in the velocity space. This process can clearly also be studied in the coordinate space, where it describes an \emph{anomalous diffusion} process. Such processes have long been observed and studied in various fields of  condensed and soft matter physics~\cite{BG90}.

Knowledge of the \emph{master equation} that is associated with a given Fokker-Planck partial differential equation is crucial for two reasons. First, as stated above the master equation
provides a uniquely defined discretization scheme for numerical integration. Second, the master equation is physically meaningful, since it defines the process dynamics on  lattices. Hence,  the study of the lattice dynamics allows an improved understanding of the dynamics in continuous space.

For example,  consider the Fokker-Planck equation in the form (\ref{I1}) that involves two arbitrary functions $a(f)$ and $b(f)$. Numerical integration of this equation requires discretization in the spatial variable.  The resulting master equation cannot contain any other functions besides $a(f)$ and $b(f)$, that are already present in the original partial differential equation. Hence, any lattice (finite-size) dynamical effects that are lost in the continuum limit cannot be re-introduced at this level. In addition, the master equation derived from the discretization of the Fokker-Planck equation depends on the particular differencing scheme (e.g., forward differences versus central differences). To avoid these shortcomings, it is worth investigating the use of physical principles \emph{at the lattice scale} in order to univocally define the discrete kinetics and directly derive the master equation. The master equation thus obtained is clearly useful, not only for deriving the nonlinear Fokker-Planck equation (\ref{I1}) in the continuum limit, but also as the starting point for the direct numerical integration of the Fokker-Planck equation.

The present paper shows that the \emph{Kinetic Interaction Principle (KIP)} which governs the particle kinetics \cite{PhysicA2001} univocally defines a very simple lattice-based master equation, which  yields the nonlinear Fokker-Planck equation  in the continuous limit. Both the master equation and the ensuing Fokker-Planck equation involve physically motivated terms.

The paper is organized as follows. In Sect.~2,  we introduce the specific form of the Fokker-Planck equation that
we will investigate. We then recall the KIP that underlies the kinetics of a particle system, and we apply it to derive the most general nonlinear master equation in the nearest-neighbour approximation. Furthermore we express the master equation in the form of a discrete continuity equation and introduce the nonlinear particle current. In Sect.~3, we calculate the particle current in the continuous limit and  derive the respective Fokker-Planck equation (\ref{I2}). In Sect.~4 we focus on standard finite-difference-based discretization schemes of the Fokker-Planck equation, and we show that in the nonlinear case the KIP-based master equation cannot be derived from such discretization schemes. In Sect.~5, we focus on an important subclass of nonlinear kinetics, described by Eq.~(\ref{I2}), which is related to ordinary Fickian diffusion.  Finally, in Sect.~6 we present our  concluding remarks and suggest future extensions of this research to lattices in higher dimensions.

\section{Kinetic Interaction Principle and Master Equation}

\subsection{Nonlinear Fokker-Planck Kinetics in One Dimension}

In the following, we  focus on the one-dimensional nonlinear diffusive process  that is described by the equation
\begin{eqnarray}
\frac{\partial f}{\partial t}= \frac{\partial}{\partial x}
\left \{\!D\, a(f)\, b(f)\frac{\partial}{\partial x} \bigg [
\beta \, U\!+\! \ln \frac{a(f)}{b(f)} \bigg ] \! \right \}. \ \  \ \ \ \   \label{I2}
\end{eqnarray}
where $f=f(t,x)$, $U=U(x)$ and $D=D(x)$, while $x$ can indicate either the velocity or the coordinate space variable.
Eq. (\ref{I2}) represents a general form of the nonlinear Fokker-Planck equation.
Upon setting $\gamma (f)= a(f)\, b(f)$, $\Omega(f)=a(f)\, b(f)\,\frac{d}{d f}  \ln \frac{a(f)}{b(f)}$, and $A=D(x)\, \beta \,\frac{\partial U(x)}{\partial x}$, this equation can also be expressed in the equivalent but more familiar form
\begin{eqnarray}
\frac{\partial f}{\partial t}= \frac{\partial}{\partial x}
\left [ A \, \gamma(f)
+ D\, \Omega (f)\, \frac{\partial f}{\partial x} \right ], \ \  \ \ \ \   \label{I3}
\end{eqnarray}
where the first summand inside the bracket on the right-hand side yields the \emph{drift term} while the second summand yields the \emph{diffusive term}.

It is in general difficult or even impossible to explicitly solve Fokker-Planck equations that are nonlinear with respect to the probability density function, or even linear ones for general forms of the external potential function $U(x)$.
In the linear case it is possible to obtain explicit approximations for certain temporal characteristics of the Fokker-Planck equation, such as the escape time from a metastable potential function (by means of the Kramers approximation) or estimates of mean first passage times~\cite{Kramers02}.

Several special cases of the nonlinear Fokker-Planck equation (\ref{I3}) have long been known. The best known and most frequently studied case in the literature is that of \emph{diffusion in porous media}, which corresponds to $A=0$, $D=$ constant and $\Omega(f)=f^n$~\cite{Gold92}. Eq.~(\ref{I3}) (expressed in velocity space) has also been used in statistical physics, by setting $A=\beta v$, $D=$ constant and $\gamma(f)=f$. When $\Omega(f)=1\pm f$, Eq.~(\ref{I3})  describes the kinetics of bosons (positive sign) or fermions (negative sign) \cite{ClassicalBF}, whereas when $\Omega(f)=f^n$, it describes particle kinetics in non-extensive statistical mechanics \cite{Plastino}.

In the case of bosonic and fermionic statistics,  Eq. (\ref{I3}) was obtained with standard methods used in linear kinetics, i.e., by applying the \emph{Kramers-Moyal expansion} to a balance equation or alternatively starting from a master equation \cite{ClassicalBF}. The same techniques were also used in~\cite{Curado,Nobre} to derive Eq. (\ref{I3})
starting from a master equation,  in the special case where the
function $\Omega (f)$ is the sum of an arbitrary number of powers
of  the function $f$, i.e. $\Omega(f)=\sum_{i} c_i f^{\alpha_i}$.

The most general form of Eq. (\ref{I3}) involves arbitrary functions  $\gamma(f)$ and $\Omega(f)$. The study of this general form is important for  classifying the  stationary and thermodynamically stable solutions \cite{Frank99,PhysicA2001,Frank01,Chavanis}. On the other hand, the derivation of the nonlinear Fokker-Planck equation from physical principles is crucial for understanding the nature of the $\gamma(f)$ and $\Omega(f)$ functions that determine the solutions.

The nonlinear Fokker-Planck equation in the most general possible from (\ref{I2})  has been obtained by applying the \emph{Kramers-Moyal expansion} (that takes into account only transitions between near neighbors) to a balance equation based on the  (KIP)~\cite{PhysicA2001}.

Subsequently, Eq.~(\ref{I3}) was obtained in \cite{Schwammle} from a master equation that involves nearest-neighbor transitions, the rates of which are determined by two arbitrary functions ${\rm a}(f)$ and $Y(f,f ')$. The first function, ${\rm a}(f)$, depends on the particle density function $f$ and is related to the $\gamma(f)$ function through the simple relation $\gamma(f)=f{\rm a}(f)$. On the other hand,  $Y(f,f')$ depends on two distinct particle density functions, $f$ and $f'$ and is related to the $\Omega(f)$ function by means of a more complicated expression.

\subsection{The Master Equation}
Let us consider the particle kinetics in a one-dimensional lattice  gas comprising $N$ identical particles. Under the hypothesis that only transitions involving the nearest-neighbour lattice sites are allowed during the evolution of the particle system, the probability density function $f_i=f(t, x_i)$ obeys the following master equation

\begin{eqnarray}
\frac{d f_i}{\ d t}= \, \pi(t,x_{i-1}\rightarrow x_{i})- \pi(t,x_{i}\rightarrow x_{i- 1})  + \, \pi(t,x_{i+1}\rightarrow x_{i})- \pi(t,x_{i}\rightarrow x_{i+ 1})  \ \  ,   \label{II1}
\end{eqnarray}
where $\pi(t,x_i \rightarrow x_j)$ is the transition probability from  site $i$ to site $j$.

The transition probability related to the hopping of a particle from site $i$ to site $j$  depends on the nature of the interaction between the lattice and the particle, and this dependence is taken into account through the \emph{transition rates} $w_{ij}$. The transition probability can of course depend on the particle density of the starting $i$ and arrival $j$ sites.

In ref. \cite{PhysicA2001}, it was postulated that the $f_i$ and $f_{j}$ populations, of the starting and arrival sites affect the transition probability through the two arbitrary functions  $a(f_i)$, and $b(f_j)$, so that
$\pi(t,x_i \rightarrow x_j)$
assumes the following factorized form

\begin{equation}
\pi(t,x_i \rightarrow x_{j})=
w_{ij} \, \, a(f_i) \, b(f_{j}) \
\ , \label{II2}
\end{equation}
and this particular dependence on the particle population of the starting and arrival sites defines the \emph{Kinetic Interaction Principle.}

\subsection{The Kinetic interaction Principle}
Let us now consider the \emph{transition probability} from site $i$ to site $j$. The $a(f_i)$ function must satisfy the obvious condition $a(0)=0$, because in the case where the starting site is empty no particles can transit toward the arrival site. On the other hand the $b(f_j)$ function must satisfy the condition $b(0)=1$, because in the case where the arrival site is empty the transition probability cannot be influenced by the arrival site.

In linear kinetics, $a(f_i)=f_i$ and $b(f_j)=1$, is usually posed \cite{Delsanto}. In \emph{nonlinear kinetics} the $b(f_j)$ function plays an important role, because it can stimulate or inhibit the particle transition to site $j$ and can simulate collective interactions in many body physics. For instance, the expression $b(f_j)=1 - f_j$  can account for the \emph{Pauli exclusion principle} that governs a  fermionic  particle system~\cite{Huber,Richards,ClassicalBF}. Analogously, the expression $b(f_j)=1+f_j$  accounts for  \emph{bosonic systems}. Other more complicated expressions of the $b(f_j)$ factor can take into account collective interactions introduced by the \emph{Haldane generalized exclusion} principle originating from the fractal structure of the single particle Hilbert space.

Another case of nonlinear kinetics corresponding to $a(f_i)=f_{i} \, \mathrm{e}^{-u \,f_{i}}$ where $u>0$ and $b(f_j)=1$ was investigated in~\cite{dth16}, motivated by
observations of two different grain growth regimes in sintering studies~\cite{dth06}. In this type of master equation the value of $u$ controls the equilibrium state: a normal diffusive regime is obtained for  $u$ below a threshold, while an abnormal diffusion regime that exhibits the \emph{Matthew effect} of accumulated advantage is obtained at higher $u$.

If transitions are allowed only among the nearest lattice neighbors, i.e., $i \rightarrow j=i\pm1$, the transition rates $w_{ij}$ are non-zero only for nearest-neighbor transitions. The \emph{nearest-neighbor transition rates} are $w_i^{\pm}$, where $w_i^{+}$ corresponds to the transition  $i \to i+1$ and $w_i^{-}$ corresponds to the transition  $i \to i-1$, the KIP assumes the following form

\begin{equation}
\pi(t,x_i\rightarrow x_{i \pm 1})=
w^{\pm}_i \, a(f_i) \, b(f_{i\pm 1}) \
\ , \label{II3}
\end{equation}
while the \emph{master equation} (\ref{II1}) becomes

\begin{eqnarray}
\label{eq:master-nnb}
\frac{d f_i}{\ d t}= \, w^{+}_{i-1} \, a(f_{i-1}) \, b(f_{i}) -w^{-}_{i} \, a(f_{i}) \, b(f_{i- 1})    + \, w^{-}_{i+1} \, a(f_{i+1}) \, b(f_{i}) -w^{+}_{i} \, a(f_{i}) \, b(f_{i+ 1}) \ \  . \ \  \ \ \ \   \label{II4}
\end{eqnarray}

\subsection{Incoming and Outgoing Lattice Currents}
Let us express the \emph{master equation} (\ref{II4}) in the form

\begin{eqnarray}
\frac{\partial f_i}{\partial t} + \frac{ j^+_i - j^-_i }{\Delta x} = 0 , \ \  \ \ \ \   \label{II5}
\end{eqnarray}
where the \emph{outgoing}, $j_i^+$, and \emph{incoming}, $j_i^-$, currents are defined as follows

\begin{eqnarray}
&&j^+_i= \Delta x \, \big [ w^{+}_{i} \, a(f_{i}) \, b(f_{i+ 1}) -  w^{-}_{i+1} \, a(f_{i+1}) \, b(f_{i})\big ] \, , \ \  \ \ \ \   \label{II6}
\\[1ex]
&&j^-_i= \Delta x \, \big [ w^{+}_{i-1} \, a(f_{i-1}) \, b(f_{i}) -w^{-}_{i} \, a(f_{i}) \, b(f_{i- 1})\big ]  \, . \ \  \ \ \ \   \label{II7}
\end{eqnarray}

Taking into account the continuity relation between the incoming current at the node $i$ and the outgoing current at node $i-1$, that is,
\begin{eqnarray}
j^-_i=j^+_{\, i-1} , \ \  \ \ \ \   \label{II8}
\end{eqnarray}
the difference $j^+_i - j^-_i $ between the outgoing and the incoming currents
can be expressed exclusively in terms of the outgoing current (or alternatively in terms of the incoming current), according to
\begin{eqnarray}
j^+_i - j^-_i =\Delta^{\pm} j^{\,\mp}_i \, , \ \  \ \ \ \   \label{II9}
\end{eqnarray}
where the \emph{forward difference} $\Delta^+ g_i$ and the \emph{backward difference} $\Delta^{-} g_i$ of the discrete function $g_i$ are defined through
\begin{subequations}
\label{eq:spatial-dif}
\begin{eqnarray}
&&\Delta^+ g_i = g_{i+1} - g_{i}  \, \, ,  \label{II10} \\[1ex]
&&\Delta^- g_i = g_{i} - g_{i-1}  \, \, .   \label{II11}
\end{eqnarray}
\end{subequations}

\subsection{Continuity Form of the Master Equation}
Based on the definition of the incoming and outgoing currents given above, the master equation (\ref{II5}) assumes the following compact form
\begin{eqnarray}
\frac{\partial f_i}{\partial t} + \frac{ \Delta^{\pm} j^{\mp}_i}{\Delta x} = 0 \ \,. \ \  \ \ \ \   \label{II12}
\end{eqnarray}
The two forms of the master equation, as given by Eq.~(\ref{II12}), are asymmetric with respect to the currents. By employing the  symmetry $\Delta^- j^+_i=\Delta^+ j^-_i$, the master equation can be expressed in a symmetric form that involves both the outgoing $j^+_i$ and incoming $j^-_i$ currents, as well as both the forward $\Delta^+$ and backward $\Delta^-$ differences, i.e.,
\begin{eqnarray}
\frac{\partial f_i}{\partial t} + \frac{1}{2} \, \left (\frac{ \Delta^{+} j^{-}_i}{\Delta x}+ \frac{ \Delta^{-} j^{+}_i}{\Delta x}\right ) = 0 \ \ . \ \  \ \ \ \   \label{II13}
\end{eqnarray}

In order to write the latter equation in an even more symmetric form we introduce the \emph{discrete Fokker-Planck current}, $j_i$, according to the average
\begin{eqnarray}
j_i = \frac{j^+_i + j^-_i}{2}  \,  . \ \  \ \ \ \   \label{II14}
\end{eqnarray}
When this definition is employed, the incoming $j^{-}_i$ and the outgoing $j^{+}_i$ currents can be expressed in terms of the $j_i$ current
\begin{eqnarray}
j^{\pm}_i = j_i \pm  \frac{\Delta x}{2}\, \frac{j^+_i-j^-_i}{\Delta x} \, \,  , \ \  \ \ \ \   \label{II15}
\end{eqnarray}
and after taking into account the master equation (\ref{II5}), we obtain
\begin{eqnarray}
j^{\pm}_i = j_i \mp  \frac{\Delta x}{2}\, \frac{\partial f_i}{\partial t} \, \,  . \ \  \ \ \ \   \label{II16}
\end{eqnarray}
Finally, the following first- and second-order \emph{symmetric finite difference} operators are introduced
\begin{eqnarray}
\frac{ \Delta }{\Delta x} = \frac{1}{2} \, \left (\frac{ \Delta^+}{\Delta x}+ \frac{ \Delta^-}{\Delta x}\right )  \  , \ \  \ \ \ \   \label{II17}
\end{eqnarray}
\begin{eqnarray}
\frac{ \Delta^2 }{(\Delta x)^2}=  \frac{1}{\Delta x} \, \left (\frac{ \Delta^{+} }{\Delta x}- \frac{ \Delta^{-} }{\Delta x}\right ) \  , \ \  \ \ \ \   \label{II18}
\end{eqnarray}
and after taking into account Eq.~(\ref{II16}), the master equation (\ref{II13}) assumes the form
\begin{eqnarray}
\frac{\partial f_i}{\partial t} + \frac{ \Delta j_i}{\Delta x} + (\Delta x)^2 \frac{1}{4}\,\frac{\partial}{\partial t} \frac{ \Delta^2 f_i}{(\Delta x)^2}  = 0 \  . \ \  \ \ \ \   \label{II19}
\end{eqnarray}

The latter expression of the master equation is particulary suitable for obtaining the  continuous limit as  $\Delta x \rightarrow 0$.
First, we introduce the particle distribution function $f(t,x)$, according to
\begin{eqnarray}
f(t,x) = \lim_{\Delta x \rightarrow 0} \, f_i \ \ , \ \  \ \ \ \   \label{II20}
\end{eqnarray}
and the Fokker-Planck current $j(t,x)$, through
\begin{eqnarray}
j(t,x) = \lim_{\Delta x \rightarrow 0} \, j_i \ \ . \ \  \ \ \ \   \label{II21}
\end{eqnarray}
Subsequently, after taking into accounts the limits
$\Delta^+/\Delta x \rightarrow \partial /\partial x$, $\Delta^-/\Delta x \rightarrow \partial /\partial x$, $\Delta /\Delta x \rightarrow \partial /\partial x$, $\Delta^2/(\Delta x)^2 \rightarrow \partial^2 /(\partial x)^2$, both  Eq.~(\ref{II13}) and Eq.~(\ref{II19}) reduce to the continuity equation
\begin{eqnarray}
\frac{\partial f(t,x)}{\partial t} + \frac{ \partial j(t,x)}{\partial x}   = 0 \ \ . \ \  \ \ \ \   \label{II22}
\end{eqnarray}

The above \emph{continuity equation} is the nonlinear Fokker-Planck equation obtained in the continuous limit, starting from the master equation (\ref{II4}). In order to write the nonlinear Fokker-Planck equation explicitly, as last step we have to calculate the Fokker-Planck current $j(t,x)$, starting from its definition (\ref{II14}), and this will be the task of the next section.

\section{The Nonlinear Fokker-Planck Current}

\subsection{Lattice Expression of the Fokker-Planck Current}
In order to calculate the continuous limit of the discrete Fokker-Planck current defined in Eq.~(\ref{II21}), we express $j_i$,  defined through Eqs.~(\ref{II6}), (\ref{II7}) and (\ref{II14}), in terms of the functions $a(\cdot)$ and $b(\cdot)$ thus obtaining
\begin{eqnarray}
j_i= \,\frac{\Delta x}{2}\, \Big [\, w^{+}_{i} \, a(f_{i}) \, b(f_{i+ 1}) + \, w^{+}_{i-1} \, a(f_{i-1}) \, b(f_{i})&& \nonumber \\ \,-\,w^{-}_{i} \, a(f_{i}) \, b(f_{i- 1}) - w^{-}_{i+1} \, a(f_{i+1}) \, b(f_{i})\, \Big ]&& \!\!\!\!\! . \ \  \ \ \ \   \label{III1}
\end{eqnarray}

In order to express the current $j_i$ in a form that can be more efficiently used to compute the continuous limit, we first note that the transition rates  $w^{\pm}_{i\mp 1}$ can be
expanded in terms of the nearest-neighbour rates $w^{\pm}_{i}$  as follows
\begin{eqnarray}
w^{\pm}_{i \mp 1}= w^{\pm}_i \mp \frac{\Delta^{\mp} w^{\pm}_i}{\Delta x}\,\Delta x \,  . \label{III2}
\end{eqnarray}
Furthermore, we assume that the functions $a(\cdot)$ and $b(\cdot)$ are differentiable functions of $f_{i}$. Then,  the $a(f_{i+1})$ function can be approximated  using the
nearest-neighbour approximation of $f_{i+1}$ in terms of $f_{i}$ and the leading-order Taylor expansion of $a(f_{i+1})$ around $f_{i}$ which lead to
\begin{eqnarray}
a(f_{i+ 1})\!\!\!\!\!&&= a \left (f_{i}+ \frac{f_{i+ 1}-f_{i}}{\Delta x}\, \Delta x \right ) \nonumber \\ &&= a \left (f_{i}+ \frac{\Delta^+ f_{i}}{\Delta x}\, \Delta x \right ) \nonumber \\ &&\approx a(f_{i}) +  \frac{d \,a (f_{i})}{d \, f_i}\,\frac{\Delta^+ f_{i}}{\Delta x}\, \Delta x.  \, \,  \ \  \ \ \ \   \label{III3}
\end{eqnarray}
Following the same procedure, one obtains the approximate expressions for the functions $a(\cdot)$, $b(\cdot)$, at the nearest-neighbour sites of a given site $i$
\begin{eqnarray}
&&a(f_{i\pm 1})\approx a(f_{i}) \pm  \frac{d \,a (f_{i})}{d \, f_i}\,\frac{\Delta^{\pm} f_{i}}{\Delta x}\, \Delta x  \, \, , \ \  \ \ \ \  \label{III4} \\[1ex]
&&b(f_{i\pm 1})\approx b(f_{i}) \pm  \frac{d \,b (f_{i})}{d \, f_i}\,\frac{\Delta^{\pm} f_{i}}{\Delta x}\, \Delta x  \, \, . \ \  \ \ \ \   \label{III5}
\end{eqnarray}

After substitution of the above expressions for $a(f_{i\pm 1})$ and $b(f_{i\pm 1})$ in the equation Eq.~(\ref{III1}) of the  \emph{discrete Fokker-Planck current} $j_i$,  one obtains
\begin{eqnarray}
j_i \!\!\!&& \approx \! \left [\! (w^{+}_i \!- \!w^{-}_i)\Delta x - \frac{\Delta^- w^{+}_i +\Delta^+ w^{-}_i }{\Delta x} \, \frac{(\Delta x)^2}{2} \!\right ] a(f_i) b(f_{i}) \nonumber \\ &&+\,a(f_i) \frac{d \,b (f_{i})}{d \, f_i}\left [ w_i^+ \frac{\Delta^+ f_{i}}{\Delta x} + w_i^- \frac{\Delta^- f_{i}}{\Delta x} \right ] \frac{(\Delta x)^2}{2}\,  \,\,\,\,\,\,  \label{N8a} \nonumber \\ &&-\,b(f_i) \frac{d \,a (f_{i})}{d \, f_i}\left [ w_i^+ \frac{\Delta^- f_{i}}{\Delta x} + w_i^- \frac{\Delta^+ f_{i}}{\Delta x} \right ]\frac{(\Delta x)^2}{2} \, \,  . \,\,\,\,\,\,  \label{III6}
\end{eqnarray}

\subsection{Continuum Limit Expression of the Fokker-Planck Current}
The above expression for the discrete current can be simplified in the continuum limit, $\Delta x \rightarrow 0$, by taking into account that there is no difference
between the first-order forward, backward, and symmetric finite-differences, i.e.,
\begin{eqnarray}
\frac{\Delta^{\pm} g_i}{\Delta x} \approx  \frac{\Delta g_i}{\Delta x}  \, \, .  \label{III7}
\end{eqnarray}
Hence, the discrete Fokker-Planck current $j_i$ assumes the form
\begin{eqnarray}
j_i \!\!\!&& \approx \! \left [\! (w^{+}_i \!- \!w^{-}_i)\Delta x - \frac{\Delta \!\left (w^{+}_i \!+\! w^{-}_i \right )}{\Delta x} \, \frac{(\Delta x)^2}{2} \!\right ] a(f_i) b(f_{i}) \nonumber \\ &&+\,(w^{+}_i + w^{-}_i)\frac{(\Delta x)^2}{2}\left [ a(f_i) \frac{d \,b (f_{i})}{d \, f_i} - b(f_i) \frac{d \,a (f_{i})}{d \, f_i} \right ]\frac{\Delta f_{i}}{\Delta x}\, \ .  \label{III8}
\end{eqnarray}
The \emph{diffusion coefficient}, defined by means of

\begin{eqnarray}
D_i=\frac{w^{+}_i + w^{-}_i}{2} \, (\Delta x)^2 , \ \ \ \  \label{III9}
\end{eqnarray}
and the \emph{drift coefficient}, defined by means of

\begin{eqnarray}
J_i=(w^{-}_i - w^{+}_i) \Delta x \, , \ \ \ \  \label{III10}
\end{eqnarray}

\noindent are usually introduced into Fokker-Plank kinetics. Following these replacements, the discrete Fokker-Planck current given by Eq.~(\ref{III8}) assumes the equivalent form

\begin{eqnarray}
 j_i=  &-& \!\! \left (J_i + \frac{\Delta D_i}{\Delta x} \!\right ) a(f_i) b(f_{i}) - D_i\left [ b(f_i) \frac{d \,a (f_{i})}{d \, f_i} - a(f_i) \frac{d \,b (f_{i})}{d \, f_i} \right ]\frac{\Delta f_{i}}{\Delta x}\, . \ \ \ \ \ \ \ \  \label{III11}
\end{eqnarray}
Furthermore, the introduction of the \emph{kinetic energy function} $U_i$   according to

\begin{eqnarray}
\beta \,\frac{\Delta U_i}{\Delta x}= \frac{1}{D_i} \left (J_i + \frac{\Delta D_i}{\Delta x} \!\right ) , \ \  \ \ \ \   \label{III12}
\end{eqnarray}

\noindent allows expressing the current  in the following form

\begin{equation}
j_i=- D_i \, a(f_i)\, b(f_{i})\bigg [\beta \, \frac{\Delta U_i}{\Delta x} + \frac{d}{d \, f_i} \ln \frac{a(f_i)}{b(f_i)} \,\, \frac{\Delta f_{i}}{\Delta x} \bigg ] \, ,   \label{III13}
\end{equation}

\noindent or equivalently as

\begin{equation}
j_i=- D_i \, a(f_i)\, b(f_{i})\frac{\Delta }{\Delta x} \bigg [\beta \, U_i + \ln \frac{a(f_i)}{b(f_i)} \bigg ] \, .   \label{III14}
\end{equation}

The discrete current $j_i$, as given by the latter equation, is expressed in a form that permits straightforward evaluation of the $\Delta x \rightarrow 0$ continuous limit. Thus,  the nonlinear \emph{Fokker-Planck current} in the continuum limit is given by the following equation

\begin{equation}
j(t,x)=- D(t,x) \, a[f(t,x)]\, b[f(t,x)]\frac{\partial }{\partial x} \left \{ \beta \, U(t,x) + \ln \frac{a[f(t,x)]}{b[f(t,x)]} \right \} \, .   \label{III15}
\end{equation}

\section{Fokker-Planck Equation and Discretization Schemes}
If we insert the continuum limit of the Fokker-Planck $j(t,x)$ current, given by Eq.~(\ref{III15}), in the continuity equation (\ref{II22}) we obtain the \emph{nonlinear Fokker-Planck equation}

\begin{eqnarray}
\frac{\partial f}{\partial t}= \frac{\partial}{\partial x}
\left \{\!D\, a(f)\, b(f)\frac{\partial}{\partial x} \bigg [
\beta \, U\!+\! \ln \frac{a(f)}{b(f)} \bigg ] \! \right \}. \ \  \ \ \ \   \label{III16}
\end{eqnarray}
The above equation recovers the original Fokker-Planck equation given by equations~\eqref{I2} and~\eqref{I3}.
Thus, the derivations in the preceding section show how the nonlinear Fokker-Planck equation can be obtained from the
discrete Fokker-Planck current of Eq.~(\ref{II21}), which is derived from the simple, KIP-based master equation~\eqref{II4}.

In the following, we investigate the impact of the discretization of the time derivative in Fokker-Planck equation, as well as the impact of the space derivative discretization in the linear and nonlinear kinetic regimes. We also present an extension
of the one-dimensional formalism to two spatial dimensions.

We will express the Fokker-Planck equation~\eqref{III16} as

\begin{subequations}
\begin{align}
\label{eq:FP}
\frac{\partial f(t,x)}{\partial t}= & F(t,x), \;
\\
F(t,x) = & \frac{\partial G(t,x)}{\partial x},
\\
G(t,x) = & D\, a(f)\, b(f)\frac{\partial}{\partial x} \bigg [ \beta \, U\!+\! \ln \frac{a(f)}{b(f)} \bigg ].
\end{align}
\end{subequations}

There are various numerical schemes for the discretization of partial differential equations such as the above, including finite-difference schemes,
methods based on finite elements, finite-volume methods, spectral approaches, multigrid methods, etc.~\cite{higham15}. Each approach leads to a different set of equations that depends on the mathematical assumptions made.
Different discretization schemes for the one-dimensional Fokker-Planck equation are presented in~\cite{larsen85}.

It is most straightforward  to  compare the master equation~\eqref{eq:master-nnb}
with discretized Fokker-Planck equations that are obtained by means of finite-difference methods. Discretized versions of the functions $f(t,x)$ are denoted by means of $f_{i}^{n}$, where the index $i$ refers to the lattice site and the index $n$ to the time instant.

\subsection{Temporal discretization}
Three of the most common finite-difference methods lead to the following discretization schemes of~\eqref{eq:FP}

\begin{subequations}
\label{eq:FP-discrete}
\begin{align}
\frac{f_{i}^{n+1}-f_{i}^{n}}{\Delta t} & =  F_{i}^{n}, & \mbox{Forward Euler}
\\
\frac{f_{i}^{n+1}-f_{i}^{n}}{\Delta t} & =  F_{i}^{n+1}, & \mbox{Backward Euler}
\\
\frac{f_{i}^{n+1}-f_{i}^{n}}{\Delta t} & = \frac{1}{2} \left( F_{i}^{n} + F_{i}^{n+1} \right),  & \mbox{Crank–Nicolson}
\end{align}
\end{subequations}
where $\Delta t$ is the time step.

On the other hand, the discretization of the master equation~\eqref{eq:master-nnb} leads to

\begin{align}
\label{eq:master-discrete}
\frac{f_{i}^{n+1}-f_{i}^{n}}{\Delta t} & = \tilde{F}_{i}^{n},
\\
\tilde{F}_{i}^{n} & = w^{+n}_{i-1} \, a(f_{i-1}^{n}) \, b(f_{i}^{n}) -w^{-n}_{i} \, a(f_{i}^{n}) \, b(f_{i- 1}^{n})    +
\, w^{-n}_{i+1} \, a(f_{i+1}^{n}) \, b(f_{i}^{n}) - w^{+n}_{i} \, a(f_{i}^{n}) \, b(f_{i+ 1}^{n})
\end{align}
If we compare the  master equation~\eqref{eq:master-discrete} with~\eqref{eq:FP-discrete} we notice that the right-hand side of the former involves the function $F_{i}^{n}$ only at the current time step. In contrast, the backward Euler and the Crank-Nicolson schemes involve the values of the function, $F_{i}^{n+1}$, at the next time  step as well.

In the forward Euler discretization the time difference is determined by the function $F_{i}^{n}$, which is in general different from $\tilde{F}_{i}^{n}.$  The function $F_{i}^{n}$ is obtained by discretizing the spatial partial derivatives in $F(t,x)$. Using the forward spatial derivative defined in~\eqref{eq:spatial-dif}, the function $\tilde{F}_{i}^{n}$ can be shown to depend on the values of $a(\cdot), b(\cdot), U(\cdot)$ at the sites labeled by $i, i+1, i+2$. In contrast, the time difference of the discretized master equation~\eqref{eq:master-discrete}
depends on the lattices sites $i$ and $i\pm 1$.

\subsection{Linear Kinetics Regime}
The linear kinetics regime of~\eqref{I2} is obtained if $a(f)=f$ and $b(f)=1$. After recalling that $\gamma (f)= a(f)\, b(f)$, $\Omega(f)=a(f)\, b(f)\,\frac{\partial}{\partial f}  \ln \frac{a(f)}{b(f)}$, it follows that $\gamma (f)=f$ and $\Omega (f) =1$.  Hence, the linear Fokker-Planck equation derived from~\eqref{I2} assumes the form:

\begin{eqnarray}
\label{eq:linear-FP}
\frac{\partial f}{\partial t}= \frac{d A(x)}{dx} \, f + \left [ A(x) + \frac{d D(x)}{dx} \right ] \, \frac{\partial f}{\partial x} + D(x)  \, \frac{\partial ^2 f}{\partial x^2} \ \ , \ \  \ \ \ \   \label{VI1}
\end{eqnarray}
with $A=D(x)\, \beta \,\frac{\partial U(x)}{\partial x}$.

In order to discretize the above equation in space, we introduce at the place of the first and second order partial derivatives the following \emph{symmetric finite differences}:

\begin{subequations}
\begin{eqnarray}
&&\frac{\partial f}{\partial x} \approx \frac{f_{i+1}-f_{i-1}}{2 \Delta x}
 \ \ , \ \  \ \ \ \   \label{VI2}
\\[1ex]
&&\frac{\partial^2 f}{\partial x^2} \approx \frac{f_{i+1}- 2 f_i+f_{i-1}}{ (\Delta x)^2}
 \ \ , \ \  \ \ \ \   \label{VI3}
\end{eqnarray}
\end{subequations}
where $\Delta x$ is the uniform discretization step in space.

Then, the linear Fokker-Planck equation~\eqref{eq:linear-FP} transforms into the linear master equation

\begin{subequations}
\begin{eqnarray}
\label{eq:linear-FP-disc}
\frac{d f_i}{\ d t}= \, w^{+}_{i-1} \, f_{i-1}   + \, w^{-}_{i+1} \, f_{i+1} \,  -w_{i} \,f_{i}   \ \  ,  \ \  \ \ \ \   \label{VI4}
\end{eqnarray}
with the nearest-neighbor transition rates $w^{-}_{i+1}$ and $ w^{+}_{i-1}$ defined by

\begin{eqnarray}
&&w^{\mp}_{i\pm 1}=\frac{D_i}{(\Delta x)^2}\, \pm  \frac{D_{i+1}-D_{i-1}}{4 (\Delta x)^2}\,  \pm  \frac{A_i}{2 \Delta x}
 \ \ , \ \  \ \ \ \   \label{VI5} \\[1ex]
&&w_i=\frac{ 2 D_i}{(\Delta x)^2}\, - \,  \frac{A_{i+1}-A_{i-1}}{2 \Delta x}\,
 \ \ . \ \  \ \ \ \   \label{VI6}
\end{eqnarray}
\end{subequations}

Similarly, the nearest-neighbor master equation~\eqref{eq:master-nnb} is given in the linear regime by the following equation
\begin{equation}
\label{eq:master-nnb-lin}
\frac{d f_i}{\ d t}= \, w^{+}_{i-1} \, f_{i-1}   + \, w^{-}_{i+1} \, f_{i+1} \,  - \left( w_{i}^{+} + w_{i}^{-}\right)\,f_{i}.   \ \   \ \  \ \ \ \
\end{equation}

After observing that $w_i \approx w^{+}_{i}+ w^{-}_{i}$, the linear master equation~\eqref{eq:master-nnb-lin} derived by
means of the KIP is essentially equivalent to the master equation~\eqref{eq:linear-FP-disc} obtained from the discretization
 of the linear Fokker-Planck equation~\eqref{eq:linear-FP}.

We can thus conclude that in the case of linear kinetics the KIP-based master equation leads in the continuum limit to a  Fokker-Planck equation. Discretization of the Fokker-Planck equation using the centered-difference scheme yields the original master equation.

In the case of linear kinetics we can use either the forward or the backward Euler method, or the Cranck-Nicolson method, since the
coefficients of the Fokker-Planck equation do not depend on time. In addition, the linear dependence on the probability density function  ensures
that even for the more demanding Crank-Nicolson scheme, it suffices to solve a linear system.

\subsection{Nonlinear Kinetics}
In the case of nonlinear kinetics, we consider the Fokker-Planck equation as given in~\eqref{I3}. Expansion of the terms that involve the spatial derivatives lead to the following expression

\begin{eqnarray}
\frac{\partial f}{\partial t}= \frac{d A(x)}{d x}\, \gamma(f) +  \left [ A(x) \frac{d \gamma (f)}{df} + \frac{d D(x)}{d x}\, \Omega (f)\, \right ] \frac{\partial f}{\partial x}
+ \, D(x)\, \frac{d \Omega (f)}{df} \left (\frac{\partial f}{\partial x} \right )^2 + D(x) \,\Omega (f) \, \frac{\partial ^2 f}{\partial x^2} \ \ . \ \  \ \ \ \   \label{VI7}
\end{eqnarray}
Using the centered finite difference approximation of the spatial derivatives  for the functions $f(t,x)$, $D(x)$ and $A(x)$ we obtain the following master equation

\begin{subequations}
\label{eq:nlfpe-disc}
\begin{eqnarray}
\label{eq:nlfpe-disc-a}
\frac{d f_i}{\ d t}= \, \omega^{+}(f_i) \, f_{i-1} +  \theta_i (f_i) \, (f_{i-1})^2  + \, \omega^{-}(f_i) \, f_{i+1} \, +  \theta_i (f_i)\, (f_{i+1})^2 \,  -\omega(f_i) \,f_{i}  \,
     -  2\theta_i (f_i) \, f_{i+1} \,f_{i-1} \, \ \  , \ \  \ \ \ \   \label{VI8}
\end{eqnarray}
where the density-dependent coefficients $\omega^{\mp}(f_i)$, $\omega(f_i)$,  and $\theta_i (f_i)$ are given by

\begin{eqnarray}
&&\omega^{\mp} (f_i)=\frac{D_i}{(\Delta x)^2}\, \Omega(f_i) \pm \frac{D_{i+1}-D_{i-1}}{4(\Delta x)^2}\, \, \Omega(f_i) \pm \frac{A_i}{2\Delta x} \, \frac{d\gamma(f_i)}{df_i}
 \ \ , \ \  \ \ \ \   \label{VI9} \\
 &&\omega (f_i)=\frac{2 D_i}{(\Delta x)^2}\, \Omega(f_i) - \frac{A_{i+1}-A_{i-1}}{2 \Delta x} \, \frac{\gamma (f_i)}{ f_i}
 \ \ , \ \  \ \ \ \   \label{VI10} \\
&&\theta_{i} (f_i)=\frac{ D_i}{4(\Delta x)^2}\,  \frac{d\Omega(f_i)}{df_i}
 \ \ . \ \  \ \ \ \   \label{VI11}
\end{eqnarray}
\end{subequations}

The master equation~\eqref{eq:master-nnb} that is based on the  KIP with nearest-neighbor transitions represents different kinetics than the master equation~\eqref{eq:nlfpe-disc} that is obtained from the discretization of the nonlinear Fokker-Planck equation~\eqref{I3}. The kinetics described by
the KIP-based master equation is physically motivated at the microscopic level.  On the other hand, it is not straightforward to interpret in physical terms the kinetics described by~\eqref{eq:nlfpe-disc}. More precisely, the  first two terms on the right-hand side of~\eqref{eq:nlfpe-disc-a} can be viewed as representing transitions into the site $i$ from the site $i-1$. The population of the departure site enters linearly the transition rate in the first term, but the population of the arrival site $i$ can be a highly nonlinear and complicated function. In  the second term, the dependence on the departure site is also nonlinear in addition to that of the arrival site. Analogous remarks hold for the third and fourth terms which represent transitions from the site $i+1$ towards the site $i$. The last two terms represent exiting transitions from the sites $i$. However, there is no obvious physical mechanism that explains why the transition rate for particles exiting the site $i$ should be proportional to the product of the densities at both sites $i-1$ and $i+1$ (as dictated by the term $\theta_i (f_i) \, f_{i+1} \,f_{i-1} $). Clearly, the master equation~\eqref{eq:nlfpe-disc} does not have an obvious microscopic interpretation, even though in the continuum limit it yields the nonlinear Fokker-Planck equation~\eqref{I3}.

Based on the above observations, in order to integrate numerically the nonlinear Fokker-Planck equation it is recommended to start with the spatial discretization provided by the master equation~\eqref{eq:master-nnb}. For the time step,  the forward or backward Euler methods are simpler alternatives than the Crank-Nicolson method which leads to a nonlinear system of algebraic equations with  respective computational cost.  More importantly, in order to study  the statistical properties of a physical  system governed by lattice nonlinear kinetics, the master equation~\eqref{eq:master-nnb} is clearly the starting point, independently of the method used to integrate it (which is not the focus of the current study).

How should we understand the difference between the master equation~\eqref{eq:master-nnb} and that produced by the discretization of  the nonlinear Fokker-Planck equation~\eqref{I3},  which after all is obtained from the initial master equation~\eqref{eq:master-nnb}  in the continuum limit? The key is that in process of taking the continuum limit  information is lost. The master equation~\eqref{eq:master-nnb} contains the density functions $f_{i}$, $f_{i-1}$, $f_{i+1}$ of three sites $i, i\pm1$ through the set of functions
$I_{M}=\{  a(f_i), b(f_i), a(f_{i-1}), b(f_{i-1}), a(f_{i+1}), b(f_{i+1})  \}$. The nonlinear Fokker-Planck equation~\eqref{I3} on the other hand, contains the function $f(x)$ through its dependence on  $a(f)$, $b(f)$ and the partial derivatives $\partial f/\partial x$, $\partial^2 f/\partial x^2$.
Clearly the set $I_{M}$ that is used in the master equation~\eqref{eq:master-nnb} contains more information than the set $I_{FP}$ = $\{$ $a(f)$, $b(f)$, $\partial f/\partial x$ and $\partial^2 f/\partial x^2$  $\}$.

Due to this loss of information in the transition from the discrete (lattice-based) to the continuum model, discretization of the latter
generates the master equation~\eqref{eq:nlfpe-disc}, which is radically different from
the master equation~\eqref{eq:master-nnb}, although both master equations yield the same Fokker-Planck equation in the continuum
limit. It is important to stress that more than one master equations, including~\eqref{eq:master-nnb} imposed
by the KIP,  yield the same Fokker-Planck equation in the continuum limit. In contrast, applying the standard discretization rules to the Fokker-Planck equation, the master
equations derived are not  physically motivated and they differ from the master equation generated by the KIP. Only in the case of the linear Fokker-Planck equation,
the master equation obtained  by discretization coincides with the KIP-based master equation.



\section{Drift Current and Diffusion in Nonlinear Kinetics}
\subsection{Nonlinear Drift and Fick Currents}

Let us now write the Fokker-Plank equation (\ref{III16}) in the following form

\begin{eqnarray}
\frac{\partial f}{\partial t} + \frac{\partial ( j_{\mathrm{drift}} +j_{\mathrm{Fick}}) }{\partial x} = 0 \, , \ \  \ \ \ \   \label{IV1}
\end{eqnarray}
where the nonlinear drift current $j_{\mathrm{drift}}=j_{\mathrm{drift}}(t,x)$ and the nonlinear Fick current $j_{\mathrm{Fick}}=j_{\mathrm{Fick}}(t,x)$ are defined according to

\begin{eqnarray}
j_{\mathrm{drift}}= - \beta D \frac{\partial U}{\partial x} \, \, a(f)\, b(f) , \ \  \ \ \ \   \label{IV2}
\end{eqnarray}
\begin{eqnarray}
j_{\mathrm{Fick}} = - D\, \phi (f) \, \frac{\partial f}{\partial x} \ , \ \  \ \ \ \   \label{IV3}
\end{eqnarray}
with

\begin{eqnarray}
\phi (f) =  a(f)\, b(f)\frac{d}{d f}
\ln \frac{a(f)}{b(f)} \ . \ \  \ \ \ \   \label{IV4}
\end{eqnarray}
The term $D\phi(f)$ represents the \emph{nonlinear diffusion coefficient}. If $\phi(f)=1$, then the ordinary Fick current
$j_{\mathrm{Fick}} = - D\,  \frac{\partial f}{\partial x}$ is obtained.  The ordinary Fick current appears in (i) \emph{linear kinetics} for $a(f)=f$ and $b(f)=1$, and (ii) in the classical models of \emph{boson or fermion  kinetics}, for $a(f)=f$ and $b(f)=1\pm f$ respectively.

This important result, pertaining to the ordinary diffusion that underlies the kinetics of classical bosons and fermions,  naturally leads to the question whether there exist other nonlinear kinetics which are governed by standard Fickian diffusion.

\subsection{Nonlinear Kinetics with Fickian Diffusion}
In the following, we focus on the Fokker-Planck equation  (\ref{III16}) in the case of Fickian diffusion i.e., if the $\phi(f)$ function defined in Eq. (\ref{IV4}) is subject to the  $\phi(f)=1$ condition. In nonlinear kinetics, the generalized logarithm $\Lambda(f)$ is introduced through

\begin{eqnarray}
\Lambda (f) = \ln \frac{a(f)}{b(f)} \  . \ \  \ \ \ \   \label{IV5}
\end{eqnarray}
In terms of the generalized logarithm,  the \emph{stationary and stable} solution, $f_{s}$, of Eq.~(\ref{III16}) assumes the form

\begin{eqnarray}
\Lambda (f_s) = -\beta [U(x)-\mu]  \  , \ \  \ \ \ \   \label{IV6}
\end{eqnarray}
where $\mu$ is an arbitrary constant.

It is easy to express the $a(f)$ and $b(f)$ functions in terms of the $\Lambda (f)$ function. First, we write  Eq.~(\ref{IV5}) in the form

\begin{eqnarray}
\frac{a(f)}{b(f)} = \exp [\Lambda (f)] \ . \ \  \ \ \ \   \label{IV7}
\end{eqnarray}
Then, based on Eq.~(\ref{IV4}) and taking into account the  $\phi (f)=1$  condition, we obtain

\begin{eqnarray}
a(f)\, b(f)= \frac{1}{d \Lambda (f) / d f}
 \ . \ \  \ \ \ \   \label{IV8}
\end{eqnarray}
Finally, the solutions of the two equations above for $a(f)$ and $b(f)$ lead to

\begin{eqnarray}
&&a(f) = \left [ \frac{d \Lambda (f)}{d\,\! f} \right ]^{-1/2} \,\exp  \left [\frac{1}{2}\, \Lambda (f)\right ] \ , \ \  \ \ \ \   \label{IV9} \\[1ex]
&&b(f) = \left [ \frac{d \Lambda (f)}{d \, \!f}\right ]^{-1/2}\,\exp \, \left [ - \frac{1}{2}\, \Lambda (f)\right ] \ . \label{IV10}
\end{eqnarray}

The nonlinear Fokker-Planck kinetics described by Eq.~(\ref{III16}), when $a(f)$ and $b(f)$ are given by Eq.~(\ref{IV9}) and (\ref{IV10}), simplifies   to

\begin{eqnarray}
\frac{\partial f}{\partial t}= \frac{\partial}{\partial x}
\left [D\,\beta \frac{\partial U}{\partial x} \, \gamma (f) + D \frac{\partial  f}{\partial x}  \! \right ] \, , \ \  \ \ \ \   \label{IV11}
\end{eqnarray}
where the function $\gamma(f)$ represents the inverse of the derivative of the generalized logarithm, given by

\begin{eqnarray}
\frac{1}{\gamma (f)} = \frac{d \Lambda (f)}{d f}  \ . \ \  \ \ \ \   \label{IV12}
\end{eqnarray}
The Eq.~\eqref{IV11} describes a process that undergoes standard Fickian diffusion dominated by a nonlinear drift.
The latter imposes  the non-standard form given by Eq.~(\ref{IV6}) to the stationary distribution function.

\subsection{The Case of Kappa Kinetics}
As a working example, let us briefly consider the $\kappa$-kinetics \cite{PhysicA2001}, where the generalized logarithm is given by

\begin{eqnarray}
\Lambda(f)=\ln_{\kappa}(f)  \ \ , \label{IV14}
\end{eqnarray}
$\ln_{\kappa}(f)$ being the $\kappa$-logarithm defined by means of the equation

\begin{eqnarray}
\ln_{\kappa}(f)=\frac{f^{\kappa}- f^{-\kappa}}{2 \kappa}  \ . \ \  \ \ \ \   \label{IV15-1}
\end{eqnarray}
The  inverse of the logarithm function, i.e., the $\kappa$-exponential, is given by

\begin{eqnarray}
\exp_{\kappa} (x) = \left ( \sqrt{1+ \kappa ^2 x^2} +\kappa x \right )^{1/\kappa}  \  . \ \  \ \ \ \   \label{IV15-2}
\end{eqnarray}
In this case the $a(f)$ and $b(f)$ functions assume the expressions

\begin{eqnarray}
&&a(f) = \sqrt {\frac{2 f}{f^{\kappa}+f^{-\kappa}}}\, \exp \left [\frac{f^{\kappa}- f^{-\kappa}}{4 \kappa}  \right ] \ , \ \  \ \ \ \   \label{IV16} \\
&&b(f) = \sqrt {\frac{2 f}{f^{\kappa}+f^{-\kappa}} }\, \exp \left [- \frac{f^{\kappa}- f^{-\kappa}}{4 \kappa}  \right ] \ , \label{IV17}
\end{eqnarray}
while the respective Fokker-Planck equation becomes

\begin{eqnarray}
\label{eq:nlfp}
\frac{\partial f}{\partial t}= \frac{\partial}{\partial x}
\left ( D \beta \frac{\partial U}{\partial x} \, \frac{2 f}{f^{\kappa}+f^{-\kappa}} + D \frac{\partial  f}{\partial x} \right ) \, . \ \  \ \ \ \   \label{IV18}
\end{eqnarray}

It is remarkable that the latter equation is different from the one proposed in \cite{PhysicA2001}. Both equations are correct and describe two different nonlinear kinetics that admit the same stationary solution

\begin{eqnarray}
f_s = \exp_{\kappa} \left [ -\beta (U-\mu) \right ]  \  . \ \  \ \ \ \   \label{IV19}
\end{eqnarray}
The main difference between the two kinetics is that the present Fokker-Planck equation describes a process which undergoes \emph{ordinary Fickian diffusion},
second term on the right hand-side of~\eqref{eq:nlfp}, in the presence of \emph{nonlinear drift}, first term on the right hand-side of~\eqref{eq:nlfp}, while the Fokker-Planck equation of ref.~\cite{PhysicA2001} describes a process that undergoes \emph{non-Fickian diffusion} in the presence of  \emph{linear drift}.

\section{Conclusions}

Let us summarize the main results of the present study.
Starting from the Kinetic Interaction Principle introduced in \cite{PhysicA2001}, we obtained the evolution equation of a general statistical system defined on a one-dimensional  lattice. The KIP-based master equation in the continuum limit yielded the already known nonlinear-Fokker-Planck equation~(\ref{I2}), which is widely used in the literature to describe anomalous diffusion in condensed matter physics and nonconventional statistical physics. Our derivation provides a better, physical understanding of the  underlying many-body lattice dynamics at the microscopic level.

A second result is related to the possibility of directly using the KIP-based master equation as the numerical discretization scheme for the solution of the Fokker-Planck partial differential equation~(\ref{I2}). This approach resolves the ambiguity that results from different numerical discretization schemes of the Fokker-Planck equation.
The proposed  discretization scheme based on the master equation is physically motivated and
follows from the KIP that describes microscopic nonlinear dynamics. On the contrary,  discretization schemes based on finite differences follow from
mathematically considerations. In addition, the master equations obtained from  the \emph{nonlinear Fokker-Planck equation} by applying  different   discretization schemes cannot be obtained starting from the KIP. The master equation proposed in this paper and obtained by means of the KIP is thus uniquely defined by microscopic interactions.

Finally,  after noting that the Fokker-Planck current expresses the sum of two distinct contributions, i.e., the \emph{nonlinear drift current} and the \emph{generalized Fick current} as shown in Eqs.~\eqref{IV1}-\eqref{IV3}, we have demonstrated that some important  nonlinear kinetics formulations  proposed in the literature  can  be successfully described by \emph{ordinary Fickian diffusion}. This can be accomplished by  introducing a nonlinear drift term that is combined with Fickian diffusion. This combination leads to the same stationary solution of the Fokker-Planck equation as an equation that involves non-Fickian diffusion and linear drift.

It is straightforward to extend the KIP-based approach for the derivation of physically inspired master equations  to lattices in two and three dimensions. For example, in the case of a two dimensional square lattice whose sites are labeled  by the integer indices $i, j$, the master equation~\eqref{eq:master-nnb} with nearest-neighbor transitions becomes

\begin{eqnarray*}
\frac{d f_{i,j}}{\ d t}= w^{+}_{i-1,j} \, a(f_{i-1,j}) \, b(f_{i,j})
-w^{-}_{i,j} \, a(f_{i,j}) \, b(f_{i- 1,j})    + \, w^{-}_{i+1,j} \,
a(f_{i+1,j}) \, b(f_{i,j}) -w^{+}_{i,j} \, a(f_{i,j}) \, b(f_{i+
1,j}) \nonumber \\
+ \, r^{+}_{i,j-1} \, a(f_{i,j-1}) \, b(f_{i,j}) -r^{-}_{i,j} \,
a(f_{i,j}) \, b(f_{i,j-1})    + \, r^{-}_{i,j+1} \, a(f_{i,j+1}) \,
b(f_{i,j}) -r^{+}_{i,j} \, a(f_{i,j}) \, b(f_{i,j+1}) \  . \nonumber \\
\end{eqnarray*}

\end{document}